\documentclass{PoS}
\usepackage{psfig,epsfig,colordvi}
\def\slashed{{/}\mskip-10.0mu}

\title{Higher loop renormalization of fermion bilinear operators}

\ShortTitle{Renormalization of fermion bilinears}

\author{\speaker{A.~Skouroupathis} \thanks{Work supported in part by the
Research Promotion Foundation of Cyprus 
(Proposal Nr: $\rm ENI\Sigma X$/0506/17).}~~  and H.~Panagopoulos \\
                University of Cyprus, Department of Physics, Cyprus\\
        E-mail: \email{php4as01@ucy.ac.cy}, 
                \email{haris@ucy.ac.cy}}

\abstract{
We compute the two-loop renormalization functions, in the RI\,$^\prime$ 
scheme, of local bilinear quark operators $\bar{\psi}\Gamma\psi$, where 
$\Gamma$ denotes the Scalar and Pseudoscalar Dirac matrices, in the 
lattice formulation of QCD. We consider both the flavor non-singlet and 
singlet operators; the latter, in the scalar case, leads directly to the
two-loop fermion mass renormalization, $Z_m$.

As a prerequisite for the above, we also compute the quark field 
renormalization, $Z_{\psi}$, up to two loops.

We use the clover action for fermions and the Wilson action for gluons. 
Our results are given as a polynomial in $c_{SW}$, in terms of both the
renormalized and bare coupling constant, in the renormalized Feynman gauge. 
We also confirm the 1-loop renormalization functions, for generic gauge.

A longer write-up of the present work, including the conversion of our 
results to the $\overline{MS}$ scheme and a generalization to arbitrary 
fermion representations, can be found in \Blue{\tt arXiv:0707.2906}.
}

\FullConference{The XXV International Symposium on Lattice Field Theory\\
		 July 30-4 August 2007\\
		 Regensburg, Germany}

\begin{document}

\section{Introduction}

Studies of hadronic properties using the lattice formulation of QCD rely on
the computation of matrix elements and correlation functions of composite 
operators, made out of quark fields. A whole variety of such operators has
been considered and studied in numerical simulations, including local and
extended bilinears, and four-fermi operators. 
A proper renormalization of these operators is most often indispensable for 
the extraction of physical results from the lattice.

In this work we study the renormalization of fermion bilinears 
${\cal O}=\bar{\psi}\Gamma\psi$ on the lattice, where $\Gamma =
\hat{1},\,\gamma_5$. 
We consider both flavor singlet and nonsinglet operators.
The cases $\Gamma=\gamma_{\mu},\,\gamma_5\,\gamma_{\mu},\,\gamma_5\,\sigma_{\mu\,\nu}$, 
will be presented in a sequel to this work.
In order to obtain the  renormalization functions of 
fermion bilinears we also compute the quark 
field renormalization, $Z_{\psi}$, as a prerequisite. 

We employ the standard Wilson action for gluons and clover-improved
Wilson fermions. The number of quark flavors $N_f$, the number 
of colors $N_c$ and the clover coefficient $c_{{\rm SW}}$ are kept as 
free parameters.
Our two-loop calculations have been performed in the bare and in 
the renormalized Feynman gauge. For 1-loop quantities, the 
gauge parameter is allowed to take arbitrary values.

The main results presented in this work are the following 
2-loop bare Green's functions (amputated, one-particle irreducible (1PI)): 
$\bullet$ {\bf Fermion self-energy:} $\Sigma^L_{\psi}(q,a_{_{\rm L}})$,  
$\bullet$ {\bf 2-pt function of the scalar} $\bar{\psi}\psi:$ 
$\Sigma^L_S(q a_{_{\rm L}})$, $\bullet$ {\bf 2-pt function of the pseudoscalar} 
$\bar{\psi}\gamma_5\psi:$ $\Sigma^L_P(q a_{_{\rm L}})$ 
($a_{_{\rm L}}\,:$ lattice spacing, $q:$ external momentum)

In general, one can use bare Green's functions to construct 
$Z_{{\cal O}}^{X,Y}$, the renormalization function for 
operator ${\cal O}$, computed within a regularization $X$ 
and renormalized in a scheme $Y$. 
We employ the $RI^{\prime}$ scheme to compute the various 
renormalization functions: $Z_{\psi}^{L,RI^{\prime}}$, 
$Z_S^{L,RI^{\prime}}$, $Z_P^{L,RI^{\prime}}$. The corresponding 
quantities in the $\overline{MS}$ scheme: $Z_{\psi}^{L,\overline{MS}}$, 
$Z_S^{L,\overline{MS}}$, $Z_P^{L,\overline{MS}}$, are presented
in the longer write-up \cite{SkouPan}.

The flavor singlet scalar renormalization function is equal 
to the fermion mass multiplicative renormalization, $Z_m$, 
which is an essential ingredient in computing quark masses.

Finally, as one of several checks on our results, we construct
the 2-loop renormalized Green's functions in  $RI^{\prime}$: 
$\Sigma_{{\cal O}}^{RI^{\prime}}(q,\bar{\mu})$ (${\cal O}\equiv\psi,S,P$),
as well as their counterparts in $\overline{MS}$:
$\Sigma_{{\cal O}}^{\overline{MS}}(q,\bar{\mu})$.
The values of all these functions, computed on the lattice, 
coincide with values computed in dimensional regularization
(as can be inferred, e.g., from ~\cite{Gracey}).

The present work is the first two-loop computation of the 
renormalization of fermion bilinears on the lattice.
There have been made several attempts to estimate 
$Z_{{\cal O}}$ non-perturbatively; see Ref.\cite{SkouPan} for
a list of recent references.
Some results have also been obtained using stochastic perturbation
theory~\cite{DiRenzo}. A related computation, regarding the fermion mass
renormalization $Z_m$ with staggered fermions can be found in~\cite{Trottier}. 

\section{Formulation of the problem}
\label{Formulation}

\noindent {\bf Lattice action:} We will make use of the Wilson formulation 
of the QCD action on the
lattice, with the addition of the clover (SW)
term for fermions. 
The clover coefficient $c_{\rm SW}$ in the Langrangian 
is treated here as a free parameter;
The fermion mass $m_{\rm o}$ in the Langrangian is a free parameter. However, since we 
will be using mass independent renormalization schemes, all renormalization functions
which we will be calculating, must be evaluated at vanishing renormalized mass, that
is, when $m_{\rm o}$ is set equal to the critical value 
$m_{\rm cr}$: $m_{\rm o}\to m_{\rm cr}=0+{\cal O}(g_{\rm o}^2)$.

\noindent {\bf Definition of renormalized quantities:} As a prerequisite, we will 
need the renormalization
functions for the gluon, ghost and fermion fields ($A_\mu^a,\ c^a, \ \psi$), and for the
coupling constant $g$ and gauge parameter $\alpha$, defined as follows:
\begin{equation}
A_{\mu\,{\rm o}}^a = \sqrt{Z_A}\,A^a_{\mu}\ \ , \ \ c^a_{\rm o}=\sqrt{Z_c}\,c^a \ \ ,
\ \ \psi_{\rm o}=\sqrt{Z_{\psi}}\,\psi \ \ ,
\ \ g_{\rm o} = \mu^{\epsilon}\,Z_g\,g\ \ , \ \ \alpha_{\rm o} = Z_a^{-1}\,Z_A\,\alpha
\label{fields}     
\end{equation}
The scale $\mu$ enters the relation between $g_o$ and $g$
only in dimensional regularization ($D=4-2\epsilon$ dimensions).
We will need $Z_A,\,Z_c,\,Z_{\alpha}\,{\rm and}\,Z_g$ to 1 loop and $Z_{\psi}$ to 
2 loops. 

\noindent {\bf Definition of the $\mathbf{RI^{\prime}}$ scheme:} 
This renormalization scheme~\cite{Martinelli,Franco,Chetyrkin2000} is
more immediate for a lattice regularized theory. 
It is defined by imposing a set of normalization conditions on matrix elements
at a scale $\bar{\mu}$, where (just as in the $\overline{MS}$ scheme): 
$\bar{\mu}=\mu\,(4\pi/{\rm e}^{\gamma_{\rm E}})^{1/2}\qquad 
(\gamma_{\rm E}\ {\rm is}\ {\rm the}\ {\rm Euler}\  {\rm constant})$

In Euclidean space, the fermion self energy
$\Sigma^L_{\psi}(q,a_{_{\rm L}})=i\slashed{q}+m_{\rm o}+{\cal O}(g_o^2)$
is renormalized by:
\begin{equation}
\lim_{a_{_{\rm L}}\rightarrow 0} \left[Z_{\psi}^{L,RI^{\prime}}(a_{_{\rm L}}\bar{\mu})\,
{\rm tr}\left(\Sigma_{\psi}^L(q,a_{_{\rm L}})\,\slashed{q}\right)
/(4i\,q^2)\right]_{q^2=\bar{\mu}^2}=1
\label{ZpsiRule}
\end{equation}
The trace here is over Dirac indices; a Kronecker delta in color and
in flavor indices has been factored out of the definition of
$\Sigma_{\psi}^L$.
Similar conditions hold for $Z_c$ (extracted from the ghost self energy) 
and $Z_A$, $Z_{\alpha}$ (extracted from the gluon propagator).
We have checked explicitly that $Z^{L,RI^{\prime}}_{\alpha}=1$ 
up to one loop, in agreement with the continuum.

For consistency with the Slavnov-Taylor identities, 
$Z_g$ in the $RI^{\prime}$ scheme is defined as in the
$\overline{MS}$ scheme. In dimensional regularization ($DR$)
this is achieved by tuning
the value of $Z_g$ in such a way as to absorb only the poles
in $\epsilon$ which appear in the gluon-fermion-antifermion 1PI vertex
function $G_{A\bar{\psi}\psi}$ (together with
matching powers of $\ln(4\pi)-\gamma_{\rm E}$); this leads to a
result for $G_{A\bar{\psi}\psi}^{{\rm finite}}$ which is finite
but not unity. 
Alternatively, a similar procedure can be performed on 
the gluon-ghost-antighost vertex $G_{A\bar{c}c}$,
leading to exactly the same value for $Z_g$.

The corresponding renormalization condition can be applied on the 
lattice vertex, $G^{L}_{A\bar{\psi}\psi}$, requiring that the
renormalized expression $G^{{\rm finite}}_{A\bar{\psi}\psi}$ 
is the {\it same} as that stemming from DR,
and similarly for $G^L_{A\bar{c}c}$. 
We have calculated $Z_g^{L,RI^{\prime}}$, using either one 
of $G^{L}_{A\bar{c}c}$ or $G^{L}_{A\bar{\psi}\psi}$, 
and have verified that the same result is obtained. 

\noindent {\bf Renormalization of fermion bilinears:} 
The renormalization of lattice operators ${\cal
  O}_{\Gamma}=\bar{\psi}\,\Gamma\,\psi$ is defined by:
${\cal O}^{RI^{\prime}}_{\Gamma}=
Z^{L,RI^{\prime}}_{\Gamma}(a_{_{\rm L}}\bar\mu)\,{\cal O}_{\Gamma\,{\rm o}}$

For the scalar (S) and pseudoscalar (P) operators, 
$Z_{\Gamma}^{L,RI^{\prime}}$ can be obtained  
through the corresponding bare 2-point functions $\Sigma^L_{\Gamma}(q a_{_{\rm L}})$
(amputated, 1PI) on the lattice, in the following way: 
\begin{equation}
\lim_{a_{_{\rm L}}\rightarrow 0}\left[Z_{\psi}^{L,RI^{\prime}}\,Z_S^{L,RI^{\prime}}\,\Sigma^L_S(q a_{_{\rm L}})\right]_{q^2=\bar{\mu}^2} = \hat{1}\ \ , 
\ \ \lim_{a_{_{\rm L}}\rightarrow 0}\left[Z_{\psi}^{L,RI^{\prime}}\,Z_P^{L,RI^{\prime}}\,\Sigma^L_P(q a_{_{\rm L}})\right]_{q^2=\bar{\mu}^2} = \gamma_5
\label{ZXrulesb}
\end{equation}

Once $Z_\Gamma^{L,RI^{\prime}}$ have been calculated, one may
convert them to the $\overline{MS}$ scheme (see Ref.\cite{SkouPan}).

\section{Computation and Results}
\label{Results}

There are 28 Feynman diagrams contributing to the fermion self-energy
$\Sigma^L_{\psi}(q,a_{_{\rm L}})$ at 1 and 2 loops,
and 21 diagrams contributing to $\Sigma^L_S(q a_{_{\rm L}})$, 
$\Sigma^L_P(q a_{_{\rm L}})$.
For flavor singlet bilinears, there are 4 extra diagrams, shown in Fig.1,
in which the operator insertion occurs inside a closed fermion loop.

The evaluation and algebraic manipulation of
Feynman diagrams, leading to a code for numerical loop integration, is
performed automatically using our software for Lattice Perturbation
Theory, written in Mathematica. 
The most laborious aspect of the procedure is the extraction of
the dependence on the external momentum $q$. This is a delicate task at two
loops; for this purpose, we cast algebraic expressions (typically
involving thousands of summands) into terms which can be naively Taylor
expanded in $q$ to the required order, plus a
smaller set of terms containing superficial divergences and/or
subdivergences. The latter can be evaluated by analytical continuation to
$D>4$ dimensions, and splitting each expression into a UV-finite part
(which can thus be calculated in the continuum), and a part which is 
polynomial in $q$.

Some of the diagrams contributing to $\Sigma^L_{\psi}(q,a_{_{\rm L}})$,
$\Sigma^L_S(q a_{_{\rm L}})$ and $\Sigma^L_P(q a_{_{\rm L}})$ 
are infrared divergent when considered separately, and thus must 
be grouped together in order to give finite results, for example, 
diagrams (1-2), (3-4) in Fig.1. 
As mentioned before, all calculations should be performed at 
vanishing renormalized mass; this can be achieved by working with
massless fermion propagators, provided an appropriate fermion mass
counterterm is introduced on 1-loop diagrams.

\begin{center}
\psfig{figure=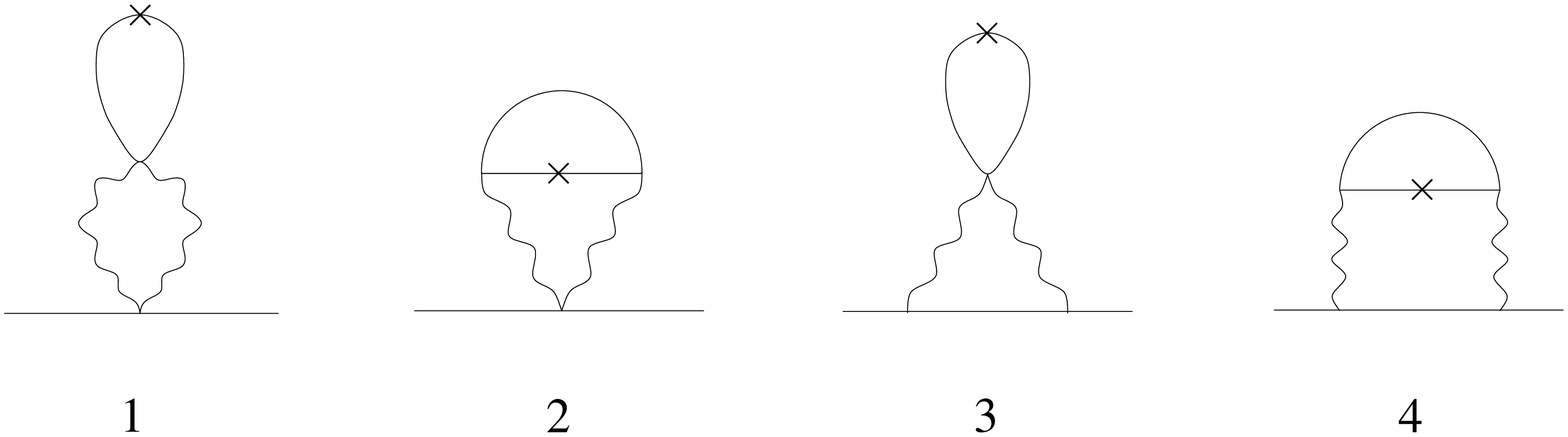,height=3.0truecm}
\vskip 1mm
{\small FIG. 1. Extra two-loop diagrams contributing to $Z_{S,\,singlet}$
A cross denotes an insertion of a flavor singlet operator.
Wavy (solid) lines represent gluons (fermions)}
\end{center}

All two-loop diagrams have been calculated in the
bare Feynman gauge ($\alpha_{\rm o}=1$). One-loop 
diagrams have been calculated 
for generic values of $\alpha_{\rm o}$; this allows 
us to convert our two-loop results
to the renormalized Feynman gauge ($\alpha_{RI^{\prime}}=1$ 
or $\alpha_{\overline{MS}}=1$). 

Numerical loop integration was carried out by our ``integrator''
program, a {\em metacode} written in Mathematica, for 
converting lengthy integrands into efficient Fortran code.
Two-loop integrals were evaluated for 
lattices of size up to $L=40$; the results were then 
extrapolated to $L\rightarrow\infty$.
The extrapolation error can
be estimated quite accurately (see, e.g. Ref.~\cite{PST}), given that
$L$-dependence of results can only span a restricted set of functional forms.

Our 1-loop results for $Z_{\psi}$, $Z_S$ and $Z_P$ confirm existing 
results~\cite{Capitani} 
(there is, however, a difference in $Z_P^{L,\overline{MS}}$, due
to an extra finite renormalization factor, $Z_5$, which is required 
in $\overline{MS}$~\cite{Larin}).

\noindent {\bf Two-loop results:} 
The evaluation of all Feynman diagrams leads directly to
the bare Green's functions $\Sigma^L_{\psi}$,
$\Sigma^L_{S}$ and $\Sigma^L_{P}$. These, in turn, can be converted to
the corresponding renormalization functions $Z_{\psi}^{L,Y}$,
$Z_{S}^{L,Y}$ and $Z_{P}^{L,Y}$ ($Y=RI^{\prime}$ or $\overline{MS}$),
via Eqs.(\ref{ZpsiRule}) and (\ref{ZXrulesb}). To
this end, we need the one-loop expression for 
$Z_A^{L,RI^{\prime}}$.
To express our results in terms of the renormalized $g$, 
we also need the one-loop expression for $Z_g^{L,RI^{\prime}}$.
Our 1-loop computation of $Z_A^{L,RI^{\prime}}$ and $Z_g^{L,RI^{\prime}}$
is in agreement with older references (see e.g. \cite{Bode}).
We present below $Z_{\psi}^{L,RI^{\prime}}$, $Z_{S}^{L,RI^{\prime}}$
and $Z_{P}^{L,RI^{\prime}}$ to two loops in the renormalized Feynman
gauge $\alpha_{RI^{\prime}}=1$.
Bare Green's functions can be easily recovered from the  
corresponding $Z$'s. The errors result from the $L\to\infty$ extrapolation. 
We employ a standard normalization for the algebra generators:
${\rm tr}(T^a\,T^b)=\delta^{ab}/2$.
($a_{\rm o}\equiv g_{\rm o}^2/16\pi^2$, $c_F\equiv (N_c^2-1)/(2\,N_c)$)
\newpage
\phantom{a}\vskip -1.5cm
\begin{eqnarray}
&\Blue{Z_{\psi}^{L,RI^{\prime}}}& = 1 + a_{\rm o}\,c_F \Blue{\bigg{[}} 
\ln(a_{_{\rm L}}^2 \bar{\mu}^2) + 11.852404288(5) - 2.248868528(3)\,c_{{\rm SW}}-1.397267102(5)\,c_{{\rm SW}}^2 \Blue{\bigg{]}} \nonumber\\
&&\hskip -0.8cm + a_{\rm o}^2\,c_F \Blue{\bigg{[}} \ln^2(a_{_{\rm L}}^2 \bar{\mu}^2) 
\left ( \frac{1}{2} c_F + \frac{2}{3} N_f -\frac{8}{3} N_c \right ) \nonumber \\ 
&&\hskip -0.8cm + \ln(a_{_{\rm L}}^2 \bar{\mu}^2) \Big{(}-6.36317446(8)\,N_f + 0.79694523(2)\,N_f\,c_{{\rm SW}} 
- 4.712691443(4)\,N_f\,c_{{\rm SW}}^2 \nonumber \\ 
&&\hskip -0.8cm + 49.83082185(5)\,c_F - 2.24886861(7)\,c_F\,c_{{\rm SW}} 
- 1.39726705(1)\,c_F\,c_{{\rm SW}}^2 
+29.03029398(4)\,N_c\Big{)} \nonumber \\ 
&&\hskip -0.8cm + N_f\,\Big{(}-7.838(2) + 1.153(1)\,c_{{\rm SW}} + 3.202(3)\,c_{{\rm SW}}^2  
+6.2477(6)\,c_{{\rm SW}}^3 + 4.0232(6)\,c_{{\rm SW}}^4 \Big{)}  \nonumber \\
&&\hskip -0.8cm + c_F\,\Big{(}505.39(1) - 58.210(9)\,c_{{\rm SW}} + 20.405(5)\,c_{{\rm SW}}^2  
+18.8431(8)\,c_{{\rm SW}}^3 + 4.2793(2)\,c_{{\rm SW}}^4 \Big{)}  \nonumber \\
&&\hskip -0.8cm + N_c\,\Big{(}-20.59(1) - 3.190(5)\,c_{{\rm SW}} -23.107(6)\,c_{{\rm SW}}^2   
-5.7234(5)\,c_{{\rm SW}}^3 - 0.7938(1)\,c_{{\rm SW}}^4 \Big{)} \Blue{\bigg{]}} 
\label{Zpsi2loopRI}
\end{eqnarray}
\vskip -6mm
\begin{eqnarray}
&\Blue{Z_S^{L,RI^{\prime}}}& = 1 + a_{\rm o}\,c_F \Blue{\bigg{[}} 3\,\ln(a_{_{\rm L}}^2 \bar{\mu}^2) -17.9524103(1) 
-7.7379159(3)\,c_{{\rm SW}} + 1.38038065(4)\,c_{{\rm SW}}^2 \Blue{\bigg{]}} \nonumber \\
&&\hskip -0.8cm + a_{\rm o}^2\,c_F \Blue{\bigg{[}} \ln^2(a_{_{\rm L}}^2 \bar{\mu}^2) \left ( \frac{9}{2} c_F + N_f -\frac{11}{2} N_c \right ) \nonumber \\
&&\hskip -0.8cm + \ln(a_{_{\rm L}}^2 \bar{\mu}^2) \Big{(}-8.1721694(5)\,N_f + 2.3908354(3)\,N_f\,c_{{\rm SW}}
-14.13807433(4)\,N_f\,c_{{\rm SW}}^2 \nonumber \\
&&\hskip -0.8cm +66.0780218(9)\,c_F - 23.213749(2)\,c_F\,c_{{\rm SW}} 
+ 4.1411425(3)\,c_F\,c_{{\rm SW}}^2 +55.7975008(9)\,N_c\Big{)} \nonumber \\
&&\hskip -0.8cm + N_f\,\Big{(}24.003(3) + 11.878(5)\,c_{{\rm SW}} + 25.59(1)\,c_{{\rm SW}}^2 
+ 22.078(3)\,c_{{\rm SW}}^3 -6.1807(8)\,c_{{\rm SW}}^4 \Big{)}  \nonumber \\
&&\hskip -0.8cm + c_F\,\Big{(}-602.35(6) + 66.80(7)\,c_{{\rm SW}} + 75.42(5)\,c_{{\rm SW}}^2  
-27.759(4)\,c_{{\rm SW}}^3 -2.688(1)\,c_{{\rm SW}}^4 \Big{)}  \nonumber \\
&&\hskip -0.8cm + N_c\,\Big{(}-38.16(4) -120.26(5)\,c_{{\rm SW}} -16.18(3)\,c_{{\rm SW}}^2 
+12.576(3)\,c_{{\rm SW}}^3 +1.0175(8)\,c_{{\rm SW}}^4 \Big{)} \Blue{\bigg{]}} 
\label{ZS2loopRI}
\end{eqnarray}
\vskip -6mm
\begin{eqnarray}
&\Blue{Z_P^{L,RI^{\prime}}}& = 1 + a_{\rm o}\,c_F \Blue{\bigg{[}} 3\,\ln(a_{_{\rm L}}^2 \bar{\mu}^2) -27.5954414(1)
+2.248868528(3)\,c_{{\rm SW}} -2.03601561(4)\,c_{{\rm SW}}^2 \Blue{\bigg{]}} \nonumber \\
&&\hskip -0.8cm + a_{\rm o}^2\,c_F \Blue{\bigg{[}} \ln^2(a_{_{\rm L}}^2 \bar{\mu}^2) \left ( \frac{9}{2} c_F + N_f -\frac{11}{2} N_c \right ) \nonumber \\
&&\hskip -0.8cm + \ln(a_{_{\rm L}}^2 \bar{\mu}^2) \Big{(}-8.1721694(4)\,N_f + 2.39083540(6)\,N_f\,c_{{\rm SW}}
-14.13807433(4)\,N_f\,c_{{\rm SW}}^2 \nonumber \\
&&\hskip -0.8cm +37.1489292(7)\,c_F +6.746606(1)\,c_F\,c_{{\rm SW}} 
-6.1080465(3)\,c_F\,c_{{\rm SW}}^2 +55.7975008(7)\,N_c\Big{)} \nonumber \\
&&\hskip -0.8cm + N_f\,\Big{(}38.231(3) -7.672(5)\,c_{{\rm SW}} + 55.32(1)\,c_{{\rm SW}}^2 
-7.049(3)\,c_{{\rm SW}}^3 +4.7469(8)\,c_{{\rm SW}}^4 \Big{)}  \nonumber \\
&&\hskip -0.8cm + c_F\,\Big{(}-876.98(4) + 85.80(2)\,c_{{\rm SW}} + 37.37(4)\,c_{{\rm SW}}^2 
+ 19.974(3)\,c_{{\rm SW}}^3 + 2.873(1)\,c_{{\rm SW}}^4 \Big{)}  \nonumber \\
&&\hskip -0.8cm + N_c\,\Big{(}-104.35(3) -38.70(2)\,c_{{\rm SW}} -13.93(3)\,c_{{\rm SW}}^2
-4.429(2)\,c_{{\rm SW}}^3 -1.2898(7)\,c_{{\rm SW}}^4 \Big{)} \Blue{\bigg{]}} 
\label{ZP2loopRI}
\end{eqnarray}

All expressions reported thus far for $Z_S$ and $Z_P$ refer to flavor
non singlet operators. In the case of $Z_P$, all diagrams
of Fig.1 vanish, so that singlet and non singlet results coincide.
$Z_S$ on the other hand, receives an additional finite contribution
(which is the same for the $\overline{MS}$ scheme as well):
\begin{eqnarray}
Z_{S,\,\rm singlet}^{L,RI^{\prime}} = Z_S^{L,RI^{\prime}} 
+ a_{\rm o}^2\,c_F N_f\,
\bigl( &-&107.76(1) + 82.27(2)\,c_{{\rm SW}} -29.727(4)\,c_{{\rm SW}}^2
\nonumber \\
&+& 3.4400(7)\,c_{{\rm SW}}^3 + 2.2758(4)\,c_{{\rm SW}}^4\bigr)
\label{ScalarSingletRI}
\end{eqnarray}

\noindent \begin{minipage}{7cm}
In Figs. 2, 3 and 4 we plot $Z_{\psi}^{L,RI^{\prime}}$, 
$Z_S^{L,RI^{\prime}}$ and 
$Z_P^{L,RI^{\prime}}$, respectively, 
as a function of $c_{\rm SW}$. For definiteness, we have set 
$N_c=3$, $\bar{\mu}=1/a_{_{\rm L}}$ and
$\beta_{\rm o}\equiv 2N_c/g_{\rm o}^2=6.0$. Our results up to two loops 
for each $Z$ are shown for both $N_f=0$ and $N_f=2$, and 
compared to the corresponding one-loop results. Furthermore, in  
the scalar case, we also present the two-loop result for the flavor
singlet operator.
\end{minipage}
\hskip 8mm
\begin{minipage}{7cm}
\begin{center}
\psfig{figure=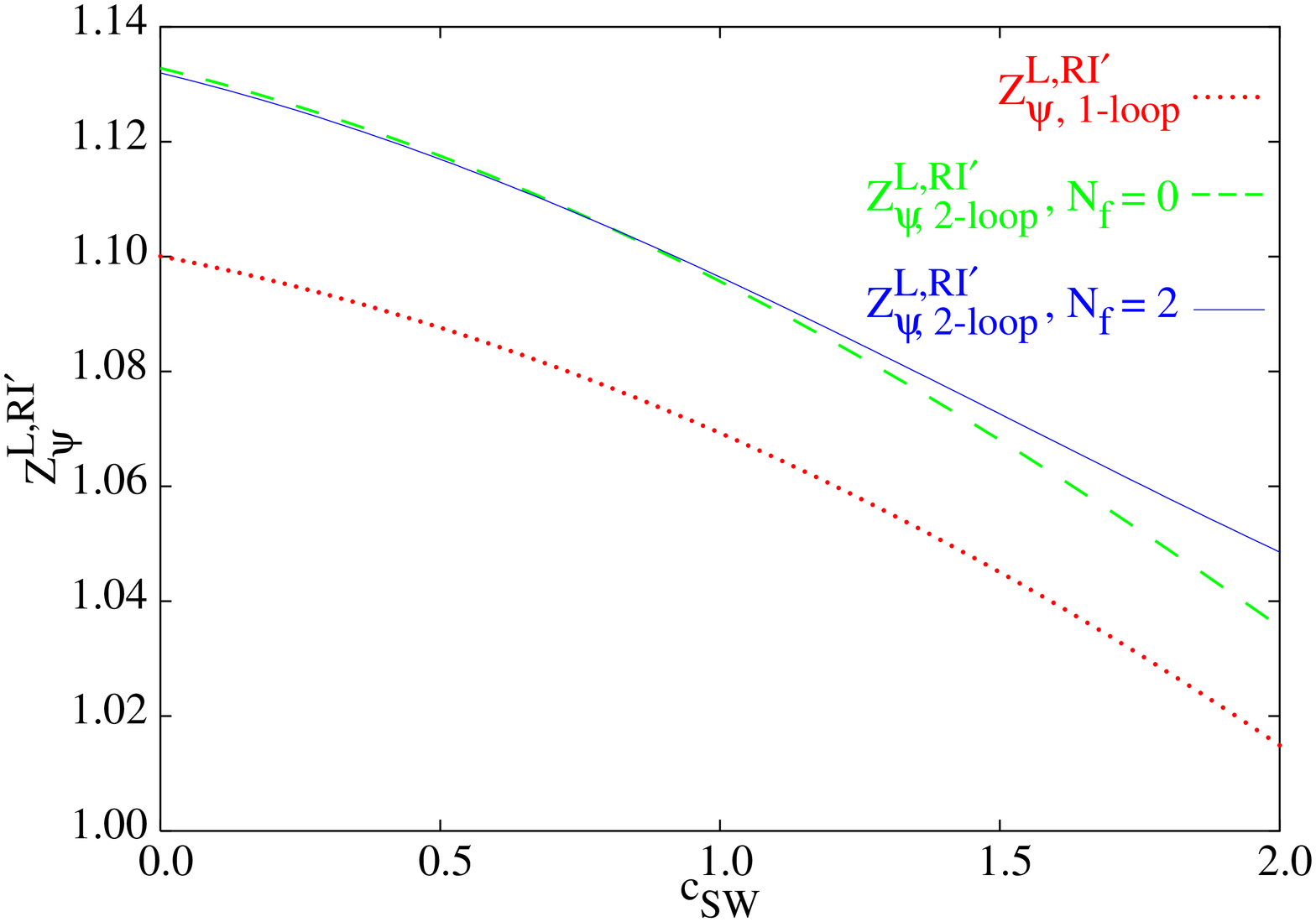,scale=0.29,angle=-90}
\vskip 4mm
{\small FIG. 2. $Z_{\psi}^{L,RI^{\prime}}(a_{_{\rm L}}\bar{\mu})$ vs. $c_{{\rm SW}}$ 
($N_c=3$, $\bar{\mu}=1/a_{_{\rm L}}$, $\beta_{\rm o}=6.0$). Results up to 2 loops are shown 
for $N_f=0$ (dashed line) and $N_f=2$ (solid line); one-loop results are plotted 
with a dotted line.}
\end{center}
\end{minipage}

\begin{center}
\begin{minipage}{7cm}
\psfig{figure=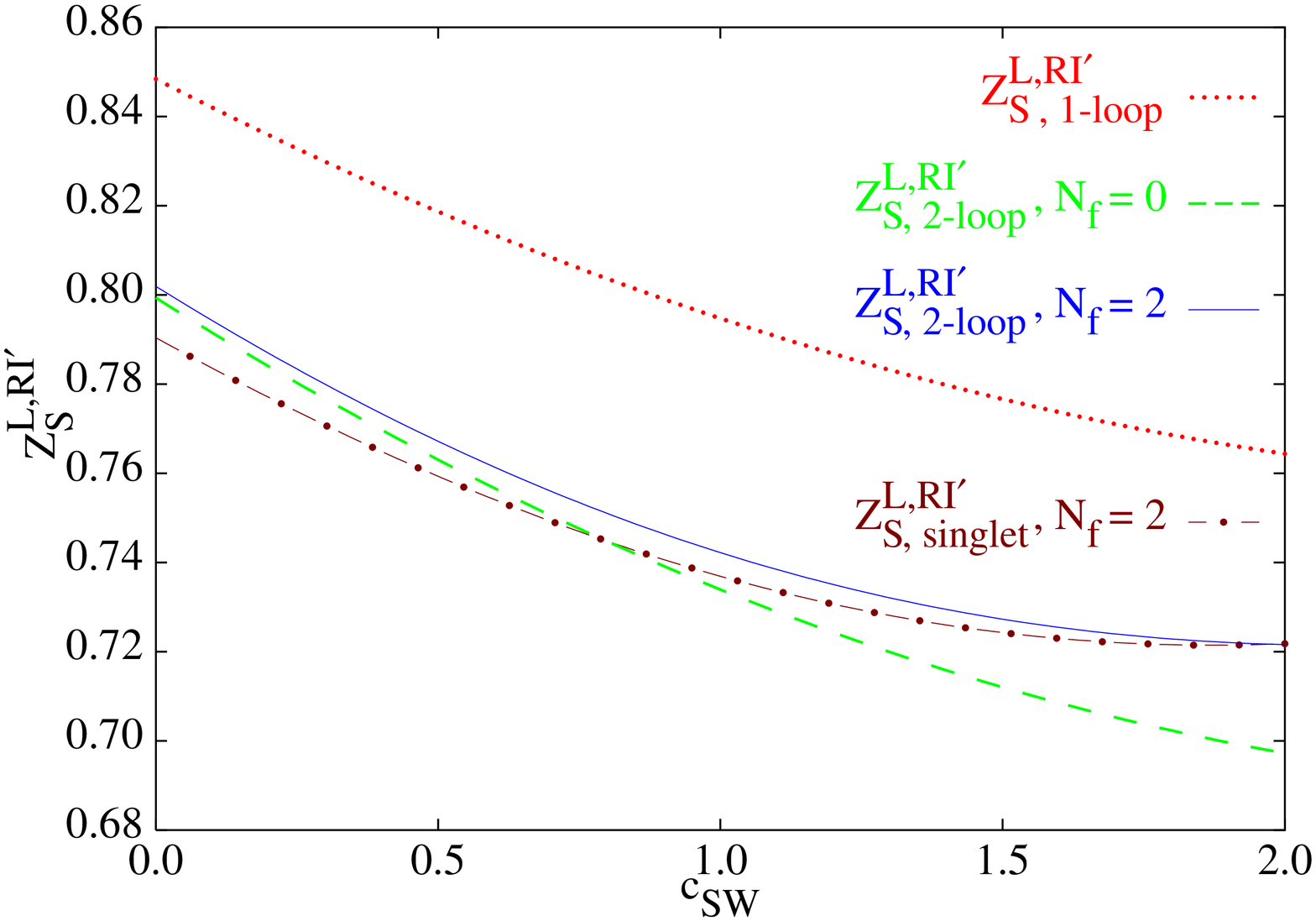,scale=0.29,angle=-90}
\vskip 4mm
\end{minipage}
\hskip 8mm
\begin{minipage}{7cm}
\psfig{figure=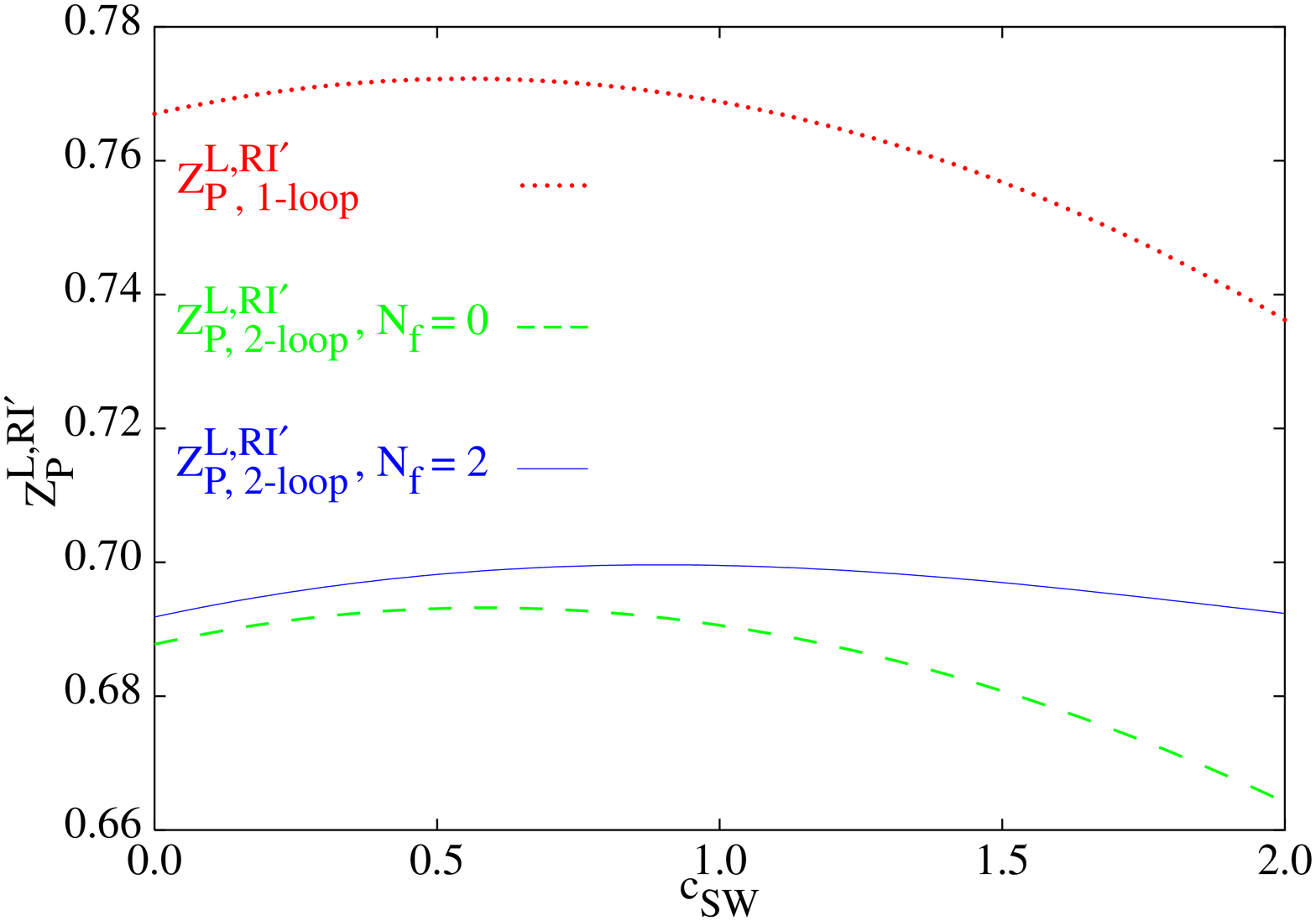,scale=0.29,angle=-90}
\vskip 4mm
\end{minipage}
\end{center}
{\small FIGS. 3, 4. $Z_S^{L,RI^{\prime}}(a_{_{\rm L}}\bar{\mu})$, 
$Z_P^{L,RI^{\prime}}(a_{_{\rm L}}\bar{\mu})$ 
vs. $c_{{\rm SW}}$ 
($N_c=3$, $\bar{\mu}=1/a_{_{\rm L}}$, $\beta_{\rm o}=6.0$). 
Results up to 2 loops, for the flavor non-singlet, 
are shown for $N_f=0$ (dashed line) and $N_f=2$ (solid line); 
2-loop results for the scalar flavor singlet, for $N_f=2$, 
are plotted with a dash-dotted line; one-loop results are 
plotted with a dotted line.}

\section{Discussion}
\label{Discussion}

As can be seen from Figs.2-4, all 2-loop
renormalization functions differ from 1-loop values in a significant
way; this difference should be taken into account in MC simulations, in order
to reduce systematic error.  At the same time, 2-loop contributions are consistently smaller
than 1-loop contributions, indicating that the (asymptotic) perturbative series are
under control. 

The dependence on the clover parameter
$c_{\rm SW}$ is also quite pronounced. In the present work, $c_{\rm SW}$ was left as
a free parameter; its optimal value, as dictated by
${\cal{O}}(a_{_{\rm L}})$ improvement, has been estimated both
non-perturbatively and perturbatively (to 1-loop) in the 
literature.

Our results regard both the flavor nonsinglet and singlet operators.
For the pseudoscalar operator, these cases coincide, just as in
dimensional regularization. The scalar, on the other hand,
receives an additional finite ($a_{_{\rm L}}\bar{\mu}$ independent)
contribution in the flavor singlet case.
$Z_{S,\,singlet}$ is seen to be equal to the fermion mass
renormalization $Z_m$\,, which is essential in the
computation of quark masses.
We note also that the scalar and pseudoscalar 
factors, necessary to convert flavor singlets from the $RI^{\prime}$
to the $\overline{MS}$ 
scheme, stay the same as their non-singlet counterparts.

The 2-loop computation of the
renormalization functions for the Vector, Axial and Tensor bilinears
is work currently in progress.

Besides the strictly local definitions of
fermion bilinears, $\bar{\psi}\Gamma\psi$, one can consider a family
of more extended operators, with the same classical continuum limit,
as dictated by ${\cal O}(a_{_{\rm L}})$ improvement. The
renormalization of these extended operators involves more Feynman
diagrams; however,
the computation is actually less cumbersome, since all additional
contributions are now free of superficial divergences. We will 
report the results of this computation in a future work.

\vskip 2mm
\noindent
{\bf Acknowledgements: } We would like to thank J. A. Gracey and S. A.
Larin for private communication regarding their continuum results.

\end{document}